\documentclass[preprint2]{aastex}

\newcommand{\beq}{\begin{equation}}
\newcommand{\beqa}{\begin{eqnarray}}
\newcommand{\eeq}{\end{equation}}
\newcommand{\eeqa}{\end{eqnarray}}

\shorttitle{Weak Lensing of Galaxy Clusters in MOND}
\shortauthors{Takahashi \& Chiba}

\begin{document}

\title{Weak Lensing of Galaxy Clusters in MOND}

\author{Ryuichi Takahashi\altaffilmark{1}
 and Takeshi Chiba\altaffilmark{2}}
\affil{\altaffilmark{1} Department of Physics and Astrophysics, 
 Nagoya University, Chikusa-ku Nagoya 464-8602, Japan}
\affil{\altaffilmark{2} Department of Physics, College of Humanities and
 Sciences, Nihon University, Tokyo 156-8550, Japan}

\begin{abstract}
We study weak gravitational lensing of galaxy clusters in terms of
 the MOND (MOdified Newtonian Dynamics) theory.
% proposed by Bekenstein (2004).
We calculate shears and convergences of background galaxies
 for three clusters (A1689, CL0024+1654, CL1358+6245) and the mean profile
 of 42 SDSS (Sloan Digital Sky Survey) clusters 
 and compare them with observational data. 
The mass profile is modeled as a sum of X-ray gas, galaxies and dark halo.
For the shear as a function of the angular radius,
 MOND predicts a shallower slope than the data irrespective of the critical 
acceleration parameter $g_0$. 
The dark halo is necessary to explain the
 data for any $g_0$ and for three interpolation functions.
If the dark halo is composed of massive neutrinos, its mass should be 
 heavier than $2$ eV.
%However we note that 
However the constraint still depends on the dark halo model
 and there are systematic uncertainties,
%, which would reduce the minimum  mass by a factor of $\sim 2$. 
 and hence the more careful study is necessary to put a stringent constraint.
%This neutrino mass is relatively high in comparison to previous 
% results ($\sim 2$ eV)
% suggested by measurement of X-ray clusters and cosmic microwave background.
% However, it is still difficult to explain a small core 
% ($30-300$ kpc) determined by the lensing data in the neutrino halo model. 

\end{abstract}


\section{Introduction}

MOND (MOdified Newtonian Dynamics) \footnote{The phrase MOND is used here to 
refer to modified Newtonian gravity models without any dark matter.} 
is a theoretical alternative                             
 to Newtonian dynamics, proposed by Milgrom (1983).
The theory itself strengthens gravitational force at large distances
 (or small accelerations)
 to explain galactic dynamics without  dark matter.
The equation of motion is changed if the acceleration is lower
 than the critical value $g_0 \simeq 1 \times 10^{-8} \mbox{cm/s}^2$. 
It is well known that this theory can explain galactic rotation curves
 with only one free parameter: the mass-to-light ratio
 (see review Sanders \& McGaugh 2002). 
There are two motivations to study such an alternative theory:
(i) General Relativity (GR) has not been tested accurately
 at much larger scale than $1$ AU
(ii) dark matter particles have not been directly
 detected and their nature still eludes us. 
Under these circumstances,
 several authors have recently studied alternative theories to GR
 (e.g. Aguirre 2003).

%While the original MOND is the modification of Newtonian dynamics,
Bekenstein (2004) recently proposed a relativistic covariant formula of
 MOND (called T$e$V$e$S) by introducing several new fields and parameters.
Following this, many authors began discussing relativistic phenomena
 such as parameterised Post-Newtonian formalism in the solar
 system \cite{b04}, gravitational lensing (e.g. Zhao et al. 2006),
 cosmic microwave background and large scale structure of the Universe
 \cite{s05,sko05,smfb06,dl06}.
In this paper, we discuss weak gravitational lensing of galaxy clusters.

Weak lensing provides an important observational method with
 which to test MOND.
This is because weak lensing probes the lens potential outside of
 the Einstein radius, $r_E \sim (M D)^{1/2}$ $\simeq 150 \mbox{kpc}
 (M/10^{14} M_\odot)^{1/2} (D/H_0^{-1})^{1/2}$, where $M$ is the lens
 mass and $D$ is the distance to the source.
%On the other hand,
The gravitational law changes outside the MOND radius,
 $r_M = (M/g_0)^{1/2}$, where the acceleration is less than $g_0$.
Since $g_0 \approx H_0/6$, the Einstein radius $r_E$ is
 a few times smaller than the MOND radius $r_M$, by a factor 2 at least,
 in the cosmological situation.
%Roughly speaking, these two radius es $r_E$ and $r_M$ have the same order of
% magnitude,  and hence
Hence, we can test the MOND-gravity regime by weak lensing.
% a low acceleration region where MOND is important.

Weak lensing is superior to X-rays as a means of probing the outer
 region of clusters.  
The lensing signal (strength of the shear) is proportional to the
 surface density $\Sigma$. 
On the other hand, the X-ray luminosity is proportional to the
 density squared $\rho^2$
 and hence X-rays can probe the inner regions of clusters. 
Hence, the outer region of the clusters can be probed with weak lensing. 

The gravitational lensing in MOND has been studied by many authors.
Before Bekenstein proposed the relativistic formula, some
 assumptions were made\footnote{For example, 
 Qin et al. (1995) assumed that the bending angle is $2$ times larger than
 that for massive particles in the limit of $m \to 0$ by analogy with GR.} 
to calculate the lensing quantities \cite{q95,mt01a,mt01b,wk01,g02}.
Just after Bekenstein's proposal, 
 Chiu et al. (2006) and Zhao et al. (2006) first studied the lensing
 in detail and tested MOND with strong lensing data of galaxies.
Zhao and his collaborators studied the gravitational lens
 statistics \cite{cz06} and investigated a non-spherical symmetric
 lens \cite{afz06}. 
Recently, Clowe et al. (2006) indicated that a merging cluster 1E 0657-558
 cannot be explained by MOND because the weak lensing mass peak
 is $8 \sigma$ spatial offset from the baryonic peak
 (= mass peak of X-ray gas).   
However, Angus et al. (2007) noted that MOND can explain the data if the
 neutrino halo is included (see also Feix et al. 2007).
Furthermore, the high observed collision velocity of the bullet clusters
 (shock velocity of $\sim 4700 \mbox{km/s}$) is more readily obtained
 in MOND than CDM \cite{am07}.

Jee et al. (2007) recently found a ring like dark matter structure
 at $\theta \sim 75^{\prime \prime}$ in
 CL0024+1654 by analysing strong and weak lensing data.
They suggested that the ring was formed by the line-of-sight collision of
 two clusters, like the bullet cluster 1E 0657-558.
Just after their discovery, Famaey et al. (2007) noted that MOND can
 easily explain the density of the ring by adding the massive neutrino of
 $2$ eV.

In this paper, we study
 three clusters (A1689, CL0024+1654, CL1358+6245) and mean profile
 of 42 SDSS (Sloan Digital Sky Survey) clusters.
We calculate shears and convergences for these clusters
 and compare them with the observational data.
We perform a $\chi^2$ fit of the data to give a constraint on the
 dark halo profile and the neutrino mass. 
%In this paper, we use MOND to mean a classical narrowly theory
% without dark matter (e.g. neutrino). 
Throughout this paper, we use the units of $c=G=1$.

\section{Basics}

We briefly review  the basics of gravitational lensing based on the
 relativistic MOND theory for a spherically symmetric lens model.
Detailed discussions are given in Bekenstein (2004) and Zhao et al. (2006).

\begin{figure}
\epsscale{.80}
\plotone{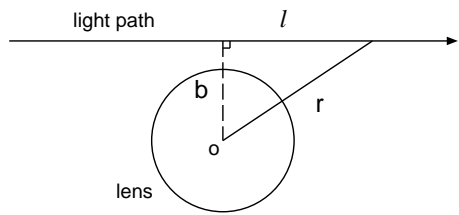}
\caption{A schematic picture of the light ray passing through the lens.
}
\label{f1}
\end{figure}

When a light ray passes through a lens with the impact parameter $b$,
 the deflection angle is
\beq
 \alpha(b) = 2 b \int_{-\infty}^{\infty} dl \frac{g(r)}{r},
\label{alpha}
\eeq
 where $l$ is the distance along the light path and $r$ is the distance
 from the lens center, $r=\sqrt{l^2+b^2}$ (see Fig.\ref{f1}). 

The gravitational force due to the lens is
\beq
 \tilde{\mu}(g/g_0) g(r)= g_N (r) = \frac{M(<r)}{r^2},
\label{gnt}
\eeq
where $g_N$ is the usual Newtonian acceleration and 
 $M(<r)$ is the lens mass enclosed inside a radius $r$.
We use a standard interpolation function\footnote{
 We also examine other interpolation functions 
 to study its dependence in section 4.}
 $\tilde{\mu}(x)=x/\sqrt{1+x^2}$
 with $g_0=1 \times 10^{-8} \mbox{cm s}^{-2}$.
Then, $\tilde{\mu}(x)=1$ (i.e. $g=g_N$) for $g \gg g_0$,
 while $\tilde{\mu}(x) = x$ (i.e. $g=\sqrt{g_N g_0}$) for $g \ll g_0$.

The lens equation is
\beq
  \theta_s = \theta - \frac{D_{LS}}{D_S} \alpha(\theta).
\label{lens-eq}
\eeq 
Here, $\theta_s$ and $\theta (=b/D_L)$ are the angular positions of the
 source and the image,
 and $D_L, D_S$ and $D_{LS}$ are the angular diameter distances between 
 the observer, the lens and the source.\footnote{We use the distance in
 the usual FRW (Friedmann-Robertson-Walker) model with $\Omega_M
 = 1-\Omega_\Lambda =0.3$ and $H_0=70 \mbox{km/s/Mpc}$.
 The distance in MOND is almost the same as that in the FRW model
 \cite{b04}.}
The shear $\gamma$ and the convergence $\kappa$ are given by,
\beq
  \gamma = \frac{1}{2} \left[ \frac{\theta_s}{\theta} -
 \frac{d\theta_s}{d\theta} \right] ; ~1-\kappa = \frac{1}{2}
 \left[ \frac{\theta_s}{\theta} + \frac{d\theta_s}{d\theta} \right]
\label{shear}
\eeq

We note that if the mass increases as $M \propto r^p$ with
 $p \geq 0$, the shear and the convergence decrease as
\beqa
  \gamma \propto \kappa \propto& \theta^{p-2}  &~~\mbox{for}~g \gg g_0,
  \nonumber  \\
                        \propto& \theta^{p/2-1}  &~~\mbox{for}~g \ll g_0,
\label{gk-prop}
\eeqa 
from Eqs.(\ref{alpha})-(\ref{shear}).
The slopes of $\gamma$ and $\kappa$ for $g \gg g_0$ are
 steeper than that for $g \ll g_0$.
This is because the gravitational force is
 proportional to $g_N^{1/2}$ for $g \ll g_0$, and hence
 the force decreases more slowly at larger distances.
Comparing the slope in Eq.(\ref{gk-prop}) with the observational data,
 we can test MOND. 

%We also give the magnification of the lensed image
%\beq
% \mu = \left| \frac{d \theta_s}{d \theta} \frac{\theta_s}{\theta}
% \right|^{-1}.
%\eeq

\section{Analysis with Cluster Data}

We calculate the shear $\gamma$ and the convergence $\kappa$ 
 based on the MOND theory for the three clusters,
 A1689, CL0024+1654, CL1358+6245, and the mean profile of 42 SDSS clusters.
The mass profiles of these clusters
 have been measured by gravitational lensing for a wider range
 of angular diameters, and hence these clusters are an appropriate
 system to investigate the angle-dependence of the shear and the
 convergence.
%In this section,
% we assume the source redshift is $z_S=1$. 

\subsection{A1689}

\begin{figure*}
\centering
  \includegraphics[width=8cm]{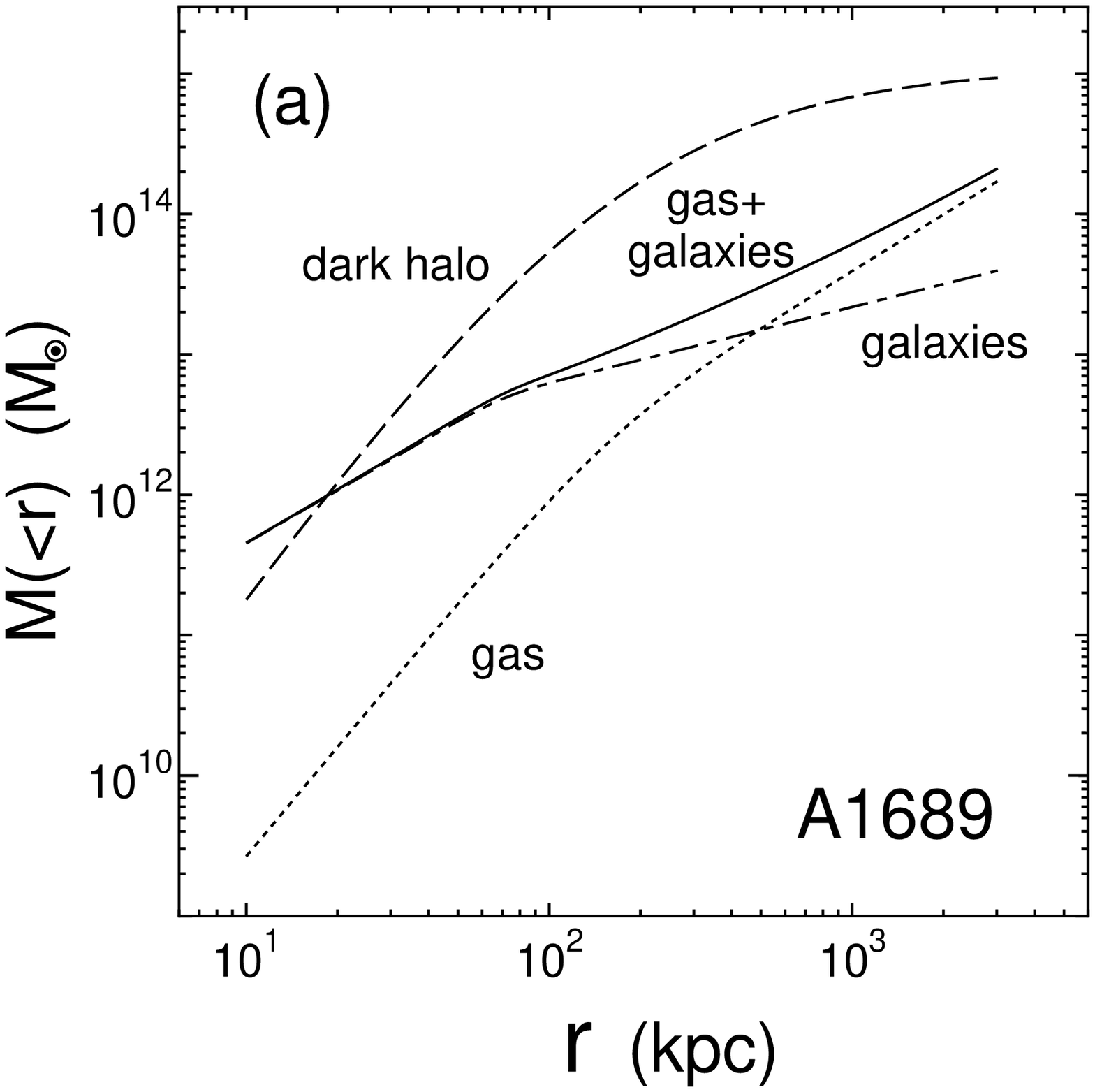}\hspace{0.5cm}\includegraphics[width=8cm]{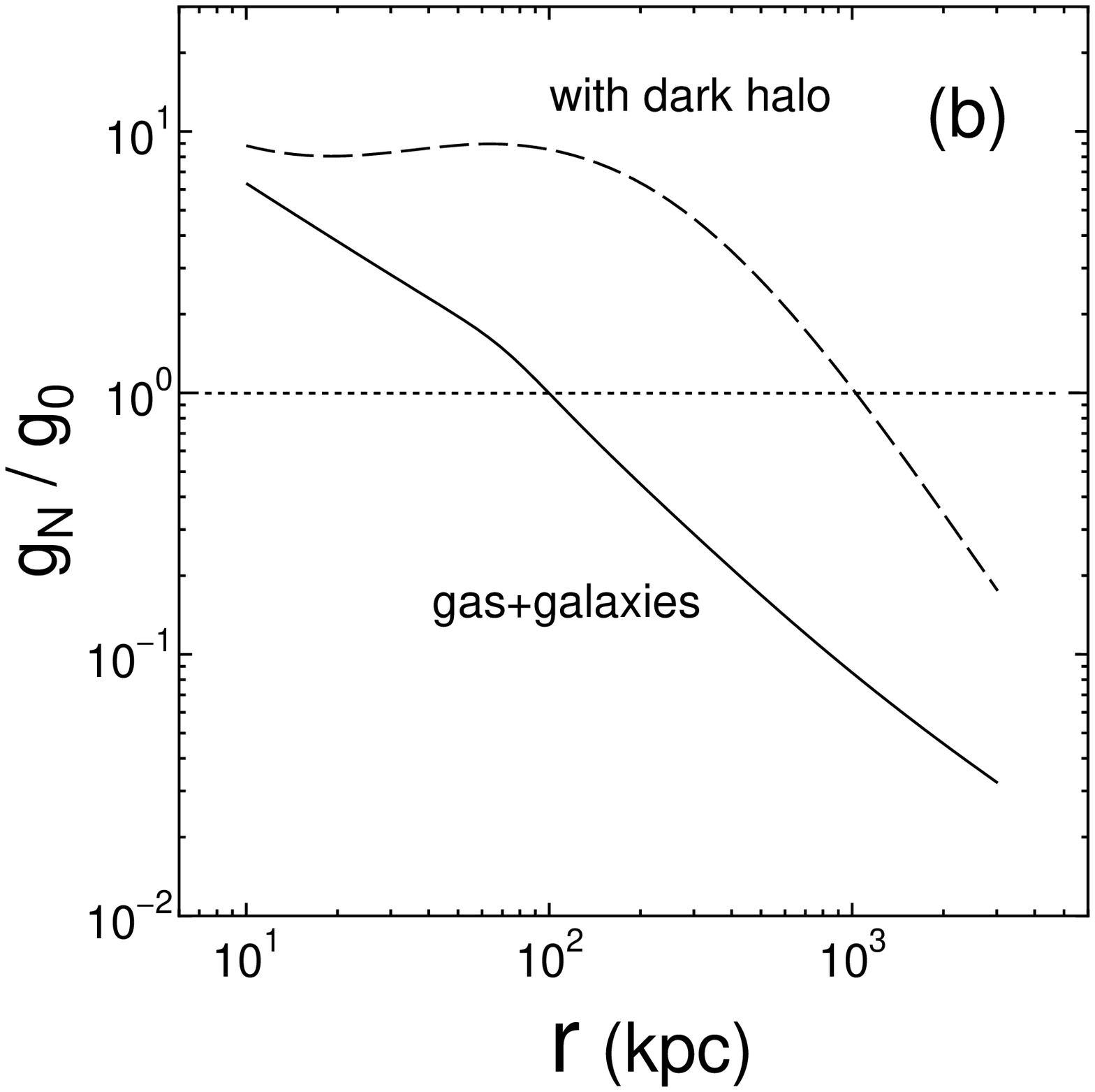}\vspace{0.5cm}
  \includegraphics[width=8cm]{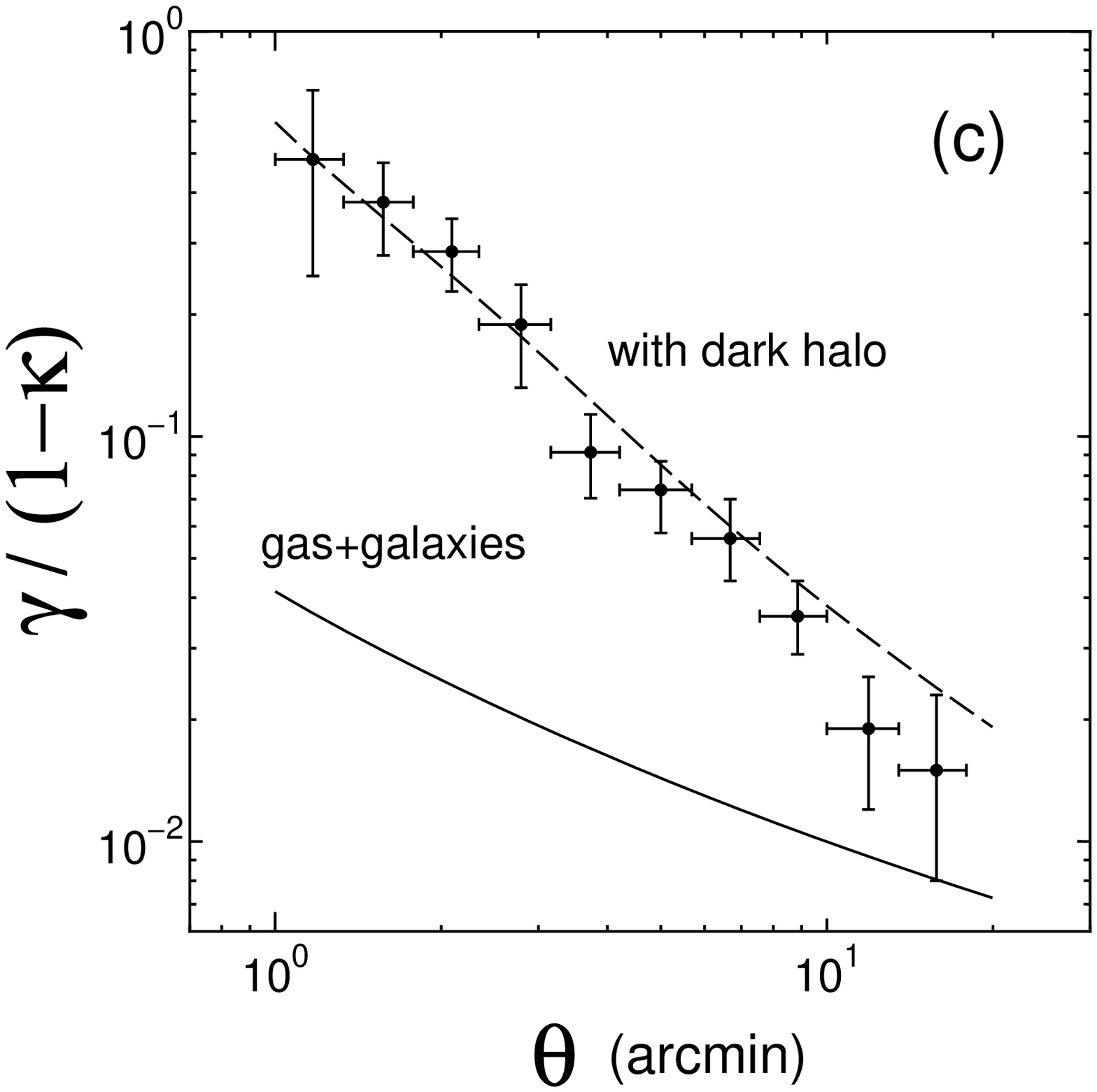}\hspace{0.5cm}\includegraphics[width=8cm]{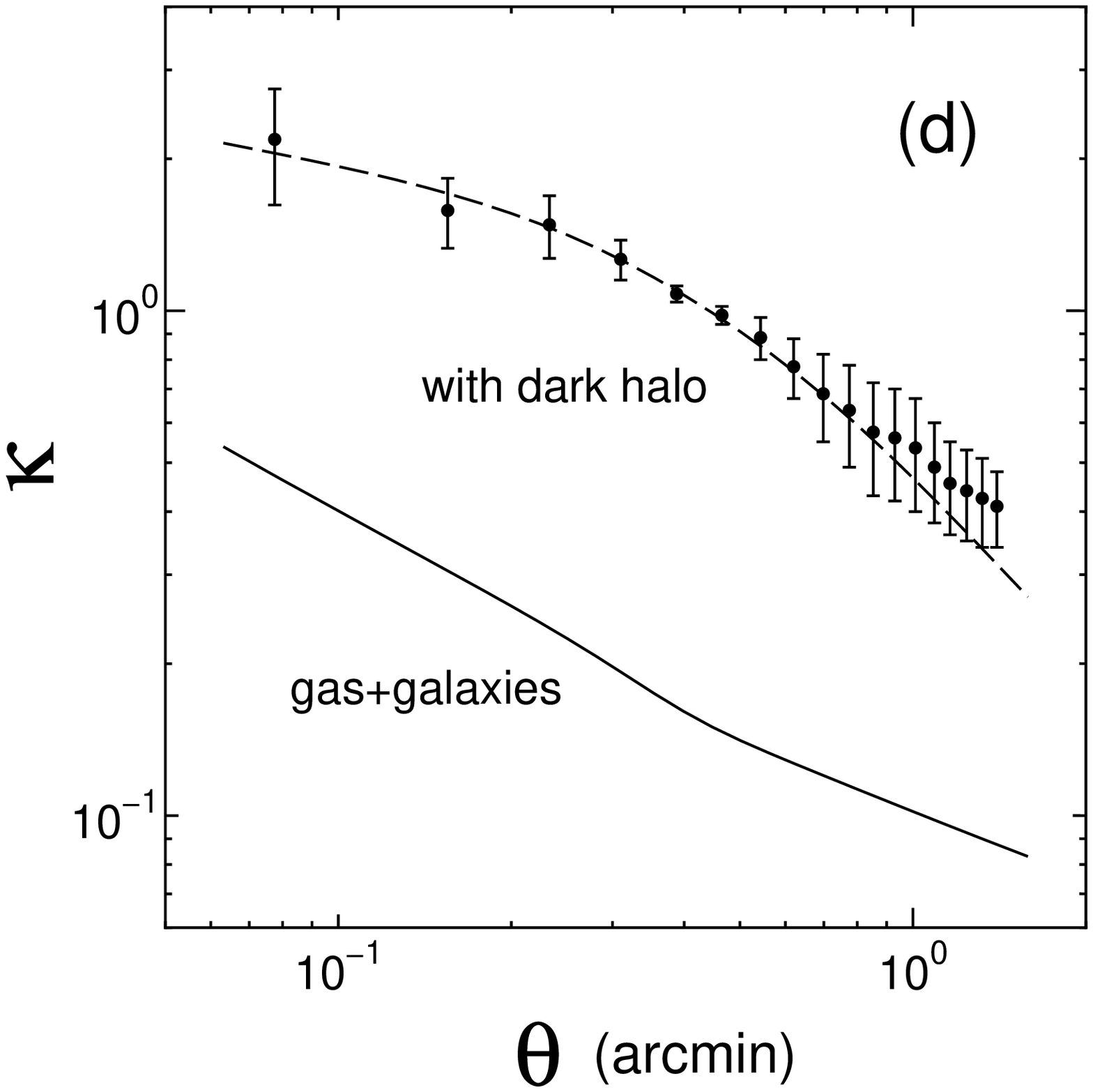}
  \caption{Results for the cluster A1689. 
The top left panel (a): The mass profiles of the gas (dotted line),
 the galaxies (dot-dashed line), the gas + galaxies (solid line),
 and the dark halo (dashed line).
The quantity $M(<r)$ is the mass enclosed within the radius $r$. 
The top right panel (b): The Newtonian gravitational acceleration $g_N$
 normalised to $g_0$ for only baryonic components (gas + galaxies)
 (solid line) and all components (dark halo is added) (dashed line).
The left bottom panel (c): The reduced shear $\gamma/(1-\kappa)$ as a 
 function of the angular radius.
The data is from Broadhurst et al. (2005a).
The solid line is the MOND prediction.
% for the standard interpolation function, $\tilde{\mu}(x)=x/\sqrt{1+x^2}$.
%On the dotted line we use the value of $g_0$, 40 times larger than the
% usual one $(=1 \times 10^{-8} \mbox{cm/s}^2)$.
On the dashed line, the dark halo is added.
The right bottom panel (d): The convergence field $\kappa$ from 
 Broadhurst (2005b). 
From panels (c) and (d), the MOND cannot explain the data unless
 the dark halo is added, because the gravitational force is too weak near
 the core.     
}
\label{a1689}
\end{figure*}

Several authors have been studying the mass profile of the rich cluster
 A1689 at $z=0.183$\footnote{$1^\prime$ corresponds to $184$kpc.}
 by strong and weak lensing, X-ray
 emission of gas, and dynamics of cluster members (e.g. 
 Limousin et al. 2006 and references therein).
The analysis of lensing data shows a small ellipticity
 ($\epsilon=0.06$ in Halkola et al. 2006) and supports the assumption of
 quasi-circular symmetry (Umetsu, Takada \& Broadhurst 2007).
%The mass estimates roughly agree with each other within less  
% the error of factor 2.  
%The mass profile of the cluster A1689\footnote{Its redshift is $z=0.183$
% and $1^\prime$ corresponds to $184$kpc.} is shown in Fig.\ref{a1689}(a).
%The hot gas mass profile was directly determined from
% X-ray observational data \cite{am04}.
Andersson \& Madejski (2004) provided the hot gas mass profile
 (40kpc $<r<$ 1Mpc) directly determined by X-ray observational data of
 the XMM-Newton telescope.
Zekser et al. (2006) gave the galaxy mass profile (20kpc $<r<$ 260kpc)
 from the surface brightness profile,
 assuming the constant mass-to-light ratio $8 M_\odot/L_\odot$
 (B-band). 
Fig.\ref{a1689}(a) shows the mass profiles of the gas (dotted),
 the galaxies (dashed), and the sum of them (solid line). 
We also show the dark halo profile which will be needed to match the
 observational data (we will discuss this later).
 
Fig.\ref{a1689}(b) shows the Newtonian gravitational acceleration $g_N$
 normalized to $g_0$.
As shown in this panel, the transition radius corresponding to
 $g_N=g_0$ (denoted by a
 horizontal dotted line) is $100$ kpc for the gas + galaxies
 and is at $1000$ kpc if the dark halo is added. 

Broadhurst et al. (2005a) measured the distortions of $6000$ red
 galaxies over $1^\prime < \theta < 20^{\prime}$ by the wide field camera,
 Suprime-Cam, of the Subaru telescope.
Panel (c) shows their results, reduced shear profile $\gamma/(1-\kappa)$.
% the convergence data $\kappa$ from strong lensing
% (Broadhurst et al. 2005b).
The mean source redshift is $z_s=1 \pm 0.1$ based on
 a photo-z estimation for deep field data.
The solid line is the MOND theoretical prediction with $z_s=1$.
The gravitational source is only baryonic component (gas + galaxies).
We note that for $\theta < 10^\prime$ the solid line is
 clearly smaller than the data.
This indicates that the gravitational force is
 too weak to explain the data.
In order to solve this discrepancy,
 we need a very high mass-to-right ratio
 $\sim 200 M_\odot/L_\odot$\footnote{The shear and the convergence
 are proportional to the mass-to-light ratio $M/L$ for $g \gg g_0$, while 
 $(M/L)^{1/2}$ for $g \ll g_0$.}.
%On the dotted line, 
% for the critical acceleration $g_0$
% we use $40$ times larger than the usual value
% ($=1 \times 10^{-8} {\mbox{cm s}}^{-2}$).
Even if the critical acceleration value $g_0$ increases,
 the discrepancy could not be resolved. 
In this case, the amplitude of the shear increases but the slope is too
 shallow to fit the data. 
%In this case, for the central region ($\theta < 4^\prime$) the theory
% can explain the data, but for a larger radius it cannot.
%This is because $g < g_0$ for the angle of $\theta > 0.5^\prime$
% $( \leftrightarrow r > 100 \mbox{kpc})$ from panel (b) and hence
% the slope is shallow from Eq.(\ref{gk-prop}).
MOND predicts shallower slope than $\gamma \propto \theta^{-1}$ for
 $g < g_0$ (since $p \geq 0$ in Eq.(\ref{gk-prop})),
 while the data in panel (c) clearly shows a steeper slope than this.
Hence MOND cannot explain the data for any mass
 model and any acceleration parameter $g_0$ in the low acceleration
 region $g < g_0$.

We comment on the dependence of the source redshift $z_s$ on the above results.
The quantities $\gamma$ and $\kappa$ depend on $z_s$ 
 through a combination of $D_{LS}/D_S$,
 $\gamma \propto \kappa \propto D_{LS}/D_S$,
 from Eqs.(\ref{lens-eq}) and (\ref{shear}).
Hence these slopes are independent of $z_s$ and the above 
 results do not change.
Furthermore the quantity $D_{LS}/D_S$ is not sensitive to $z_s$
 for relatively low lens redshift ($z=0.183$ for this cluster).

\if{}
The lensing magnification $\mu$ expands the area of sky and amplifies the
 flux of background galaxies.
The number counts of galaxies $N$ behind the cluster are changed by this
 magnification  effect \cite{btp95} :
 $N/N_0=\mu^{2.5 s-1}$ where $N_0$ is the unlensed counts and $s$ is 
 the slope of the background galaxy luminosity function. 
Panel (d) shows the magnification bias $\mu^{2.5 s-1}$
 with $s=0.22$. 
For the solid (dotted) line, the magnification bias is too strong (weak)
 to match the data.
\fi{}

Previously, Aguirre, Schaye \& Quataert (2001), Sanders (2003), and
 Pointecouteau \& Silk (2005) reached the same conclusion as ours
 by studying temperature profiles of clusters.
They indicated that the temperature data near the core is higher
 than the MOND prediction.
% and the MOND can not explain the isothermal profile.
Sanders (2003) noted that if the dark matter core were added, this
 discrepancy could be resolved.
Following the previous studies,
 we include the dark halo to explain the observational data. 
We use the dark halo with a flat core :
\beq
M(<r)=M_0 \left( \frac{r}{r+r_0} \right)^3.
\label{hp}
\eeq
Here $r_0$ is the core radius and the density steeply decreases
 with proportional to $r^{-4}$ for $r > r_0$ :
 $\rho \propto (r+r_0)^{-4}$.
We perform a $\chi^2$ fit of the data in
 order to determine the parameters $M_0$ and $r_0$.
We also use the strong lens data (convergence field $\kappa$)
 from Broadhurst (2005b) in panel (d) in order to put a strong constraint
 on the core radius.
They constructed the convergence field at $0.08^\prime < \theta <
 1.4^\prime$ from 106 multiply images of 30 background galaxies by
 Hubble Space Telescope Advanced Camera for Surveys.
The $\chi^2$ is given by $\chi^2=\sum_i (x_i-x_i^{data})^2/\sigma_{i}^2$
 where $x_i$ is the reduced shear $\gamma/(1-\kappa)$ and the
 convergence $\kappa$ at the $i$-th angle,
 $x_i^{data}$ is the data and $\sigma_i$ is the standard deviation. 
The best fitted model is $M_0=(1.1 \pm 0.1) \times 10^{15} M_\odot$
 and $r_0=174 \pm 11$ kpc. 
These relative errors are less than $10 \%$ because of combining the 
 strong and weak lensing data.  
The minimized $\chi^2$-value per degree of freedom (dof)
% \footnote{Here dof is
% the number of the data minus the free parameters.}
 is $\chi^2_{min}$/dof $=14.7/26$. 
The results are insensitive to the mass-to-light ratio.
%This best fitting model is insensitive to mass-to-light ratio. 
As shown in panels (c) and (d), this model (dashed line) fits
 the data well.
The dashed line in panel (c) is steeper than the solid line,
 since $\theta < 5^\prime$ $(\leftrightarrow r < 1000 \mbox{kpc})$ is the
 high acceleration region $g > g_0$ from panel (b) and hence the slope 
 is steeper, as can be seen in Eq.(\ref{gk-prop}).

We also try to fit the data by the other halo profiles :
 Navarro-Frenk-White (NFW) model
 $\rho \propto r(r+r_0)^{-2}$, Hernquist model $\rho \propto
 r(r+r_0)^{-3}$, isothermal (IS) with a core $\rho \propto 1/(r^2+r_0^2)$.
%These models can not fit the data.
The minimum $\chi^2_{min}$ are $29.8$ for NFW, $22.6$ for Hernquist, and
 $49.6$ for IS with core which are larger than $14.7$ for
 our model in Eq.(\ref{hp}).
This is because the convergence data favor a flat core and 
 the shear data favor a steeply decreasing profile.

\subsection{CL0024+1654}

\begin{figure*}
\centering
  \includegraphics[width=8cm]{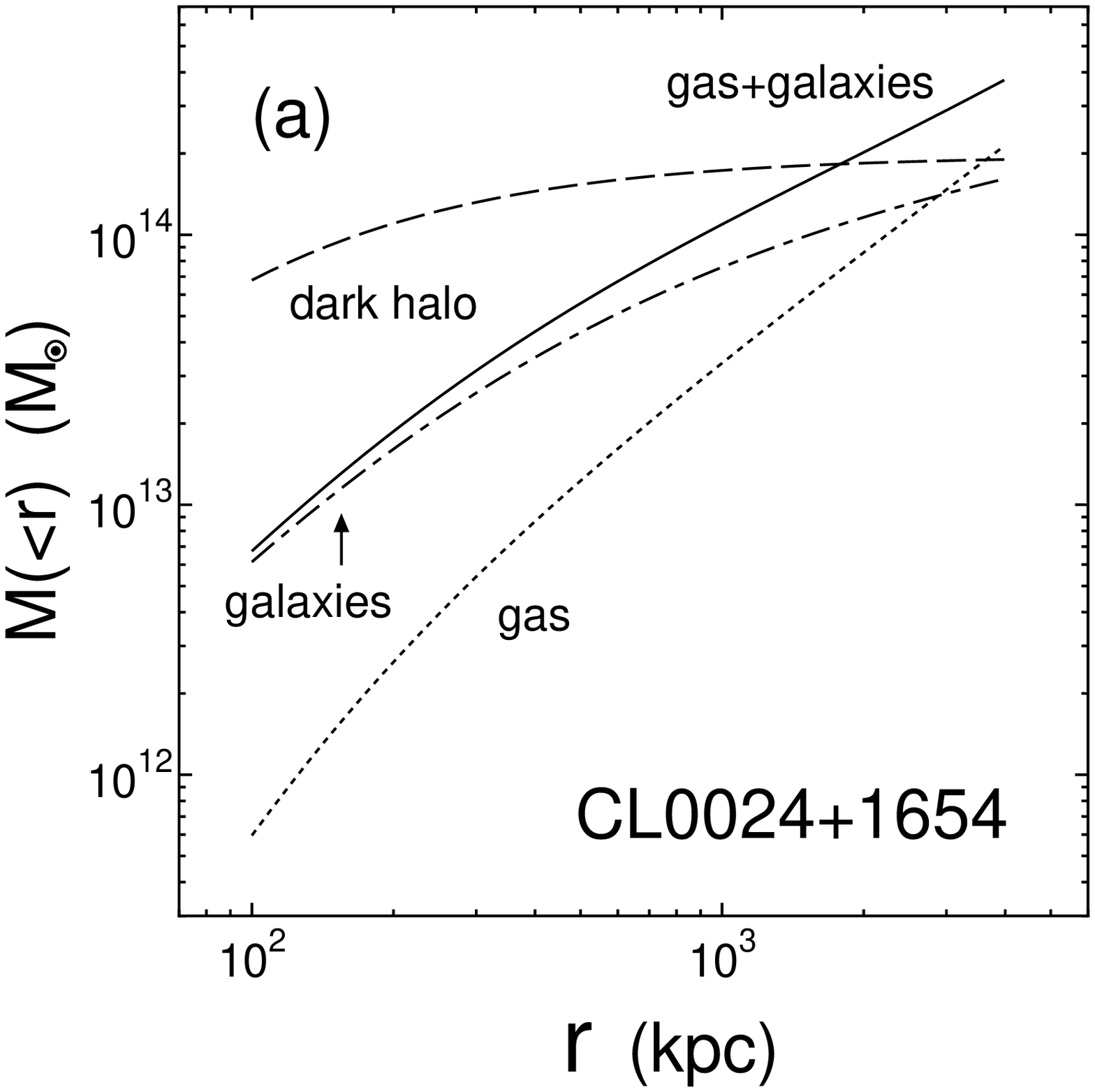}\hspace{0.5cm}\includegraphics[width=8cm]{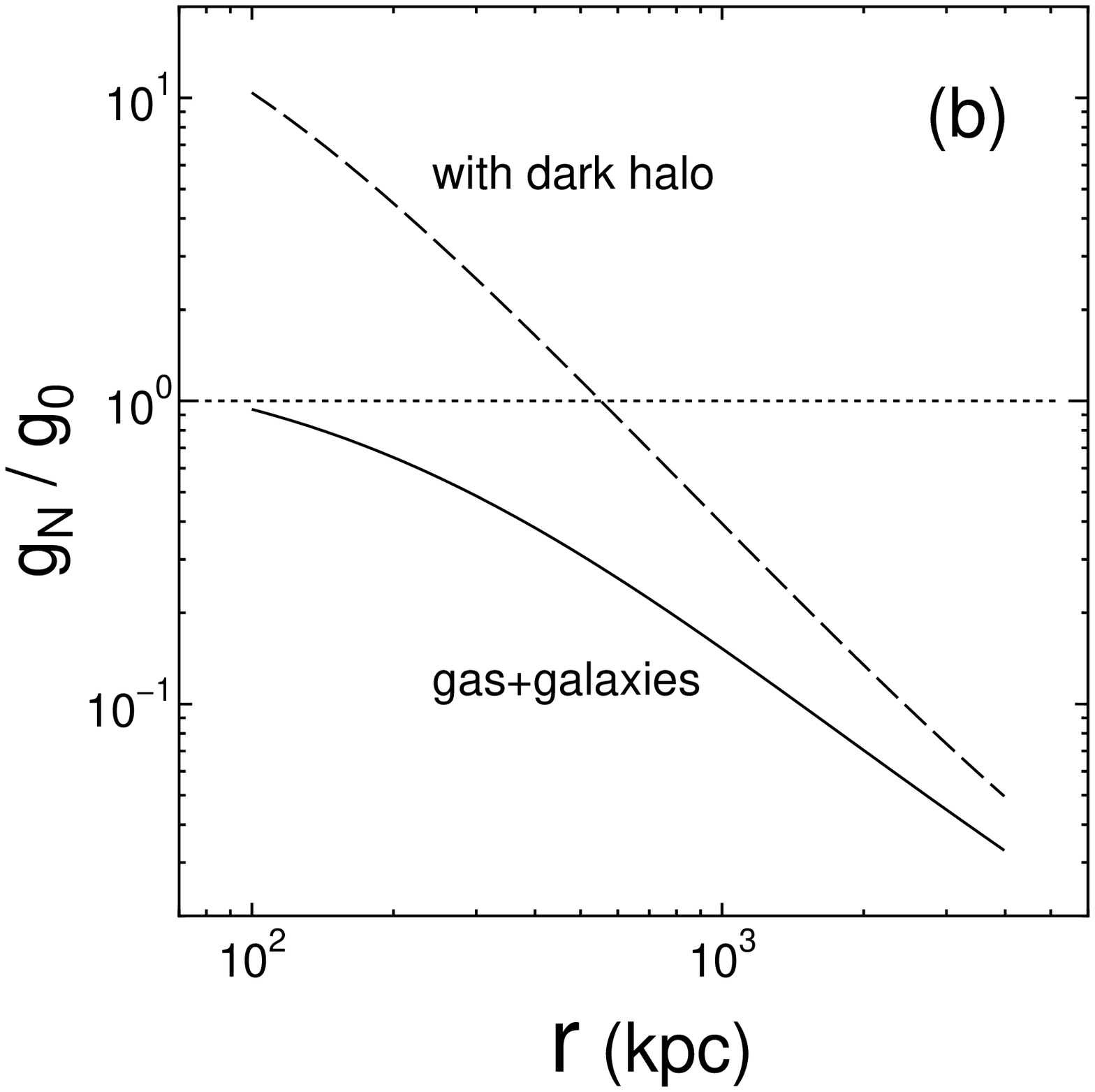}\vspace{0.5cm}
  \includegraphics[width=8cm]{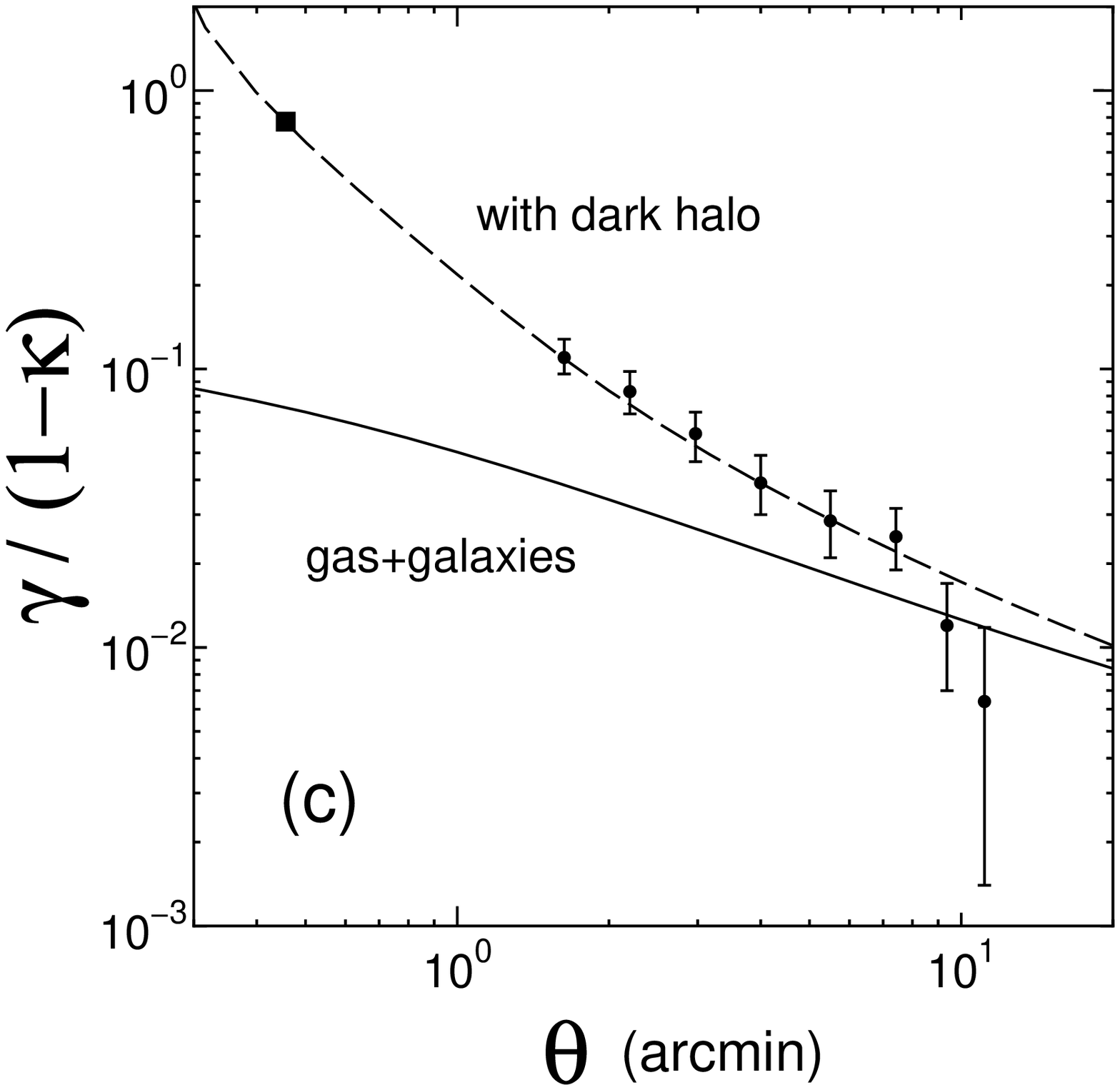}
  \caption{Same as Fig.\ref{a1689}, but for the cluster CL0024+1654. 
The black square $\blacksquare$ in panel (c) denotes the angular position of
 the Einstein ring.
}
\label{cl0024}
\end{figure*}

CL0024 is a rich cluster at $z=0.395$ ($1^\prime$ corresponds to
 $320$ kpc) with a velocity dispersion of $1200 \mbox{km s}^{-1}$ \cite{dg92}.
%There are many lensing studies about this system.
%Tyson, Kochanski \& Dell'Antonio constructed its mass profile from
% multiply images and showed the existence of the core with $35 h^{-1}$ kpc. 
%The rich cluster CL0024 is the one of the famous system to  
%Although this cluster has a second mass clump at $3^\prime$ separated
% from main clump with $30 \%$ of total cluster mass,
% we consider only main clump and assume circular symmetry for simplicity. 
Fig.\ref{cl0024} (a) shows the mass profiles of the gas determined by the
 XMM-Newton telescope \cite{z05}, 
 the galaxies with a mass-to-right ratio
 $8 M_\odot/L_\odot$ (K-band) \cite{k03}, and the dark halo. 
For the larger radius $>2$ Mpc the baryonic mass exceeds the dark
 halo mass.
This is because we extrapolate the gas profile
 (fitted by isothermal $\beta$ model for $r \lesssim 1$ Mpc)
 to the larger radius. 
Kneib et al. (2003) provided the reduced shear profile up to $10^\prime$
 measured by the Hubble Space Telescope as shown in panel (c).
%\footnote{
% After we submitted the manuscript, Jee et al. (2007) provide the shear
% profile inside of $2^\prime$ and found a dip at $\sim 75^{\prime \prime}$.
% However it may be due to some systematics, and we do not include this.}  
The mean source redshift is $z_s=1.15$.
We also use a constraint from an angular position of Einstein radius
 at $27^{\prime \prime}$ (denoted by a black square $\blacksquare$)
 based on an observation of giant arcs of distant galaxy
 at $z=1.675$ \cite{s96,b00}.
The angular position of arcs is used to set a constraint on the enclosed 
 mass within $27^{\prime \prime}$ ($= 140$ kpc). 
Similar to the previous case of A1689,
 for $\theta < 10^\prime$ the solid line is smaller than the data, and
 we need the $\sim 30 M_\odot/L_\odot$ to solve this discrepancy.
%On the dotted line, the quantity $g_0$ is $5$ times larger than the usual
% value $(=1 \times 10^{-8} \mbox{cm/s}^2)$.
The best fitted halo model is
 $M_0=(2.0 \pm 0.4) \times 10^{14} M_\odot$
 and $r_0=42 \pm 15$ kpc in Eq.(\ref{hp})
 with $\chi^2_{min}$/dof $= 5.6/7$. 
We caution that the core radius in the best fitting model
 $r_0=42 \pm 15$ is smaller
 than the inner most data point ($27^{\prime \prime}$ $= 140$ kpc), and
 hence this result has little meaning.
It only means that the enclosed mass is $\sim M_0$ inside of 140 kpc.
%Same as the previous case of A1689, 
% the dark halo is needed to fit the data.

\subsection{CL1358+6245}

The redshift of CL1358 is $z=0.33$ and $1^\prime$ corresponds to $280$ kpc.
Fig.\ref{cl0024}(a) shows the mass distribution of the gas \cite{abg02},
 the galaxies with a mass-to-right ratio $8 M_\odot/L_\odot$ (V-band)
 \cite{h98}, and the dark halo.   
Hoekstra et al. (1998) presented the reduced shear profile
 ($\theta < 4^\prime$) measured by HST
 as shown in panel (c).
They fitted the data by the isothermal sphere model with the velocity
 dispersion of $780 \pm 50$ km/s.
The solid line is the MOND prediction with
 $D_{LS}/D_S=0.62$.
% here we set $D_{}$ instead of fixing the source redshift,
% given in Hoekstra et al. (1998).
We need the $\sim 30 M_\odot/L_\odot$ to fit the data
 if we assume only baryonic components.
%On the dotted line, the quantity $g_0$ is $6$ times larger than the usual
% value. 
The best fitted model is $M_0=(6.9 \pm 2.8) \times 10^{13} M_\odot$ and
 with $r_0=69 \pm 31$ kpc with $\chi^2_{min}$/dof $= 4.5/7$.
Same as the case for CL0024, the core radius is smaller than the inner most
 data point. 
Although the discrepancy between the MOND prediction and the data is not
 so large in comparison with the previous
 cases, the dark halo model is better.

\begin{figure*}
\centering
  \includegraphics[width=8cm]{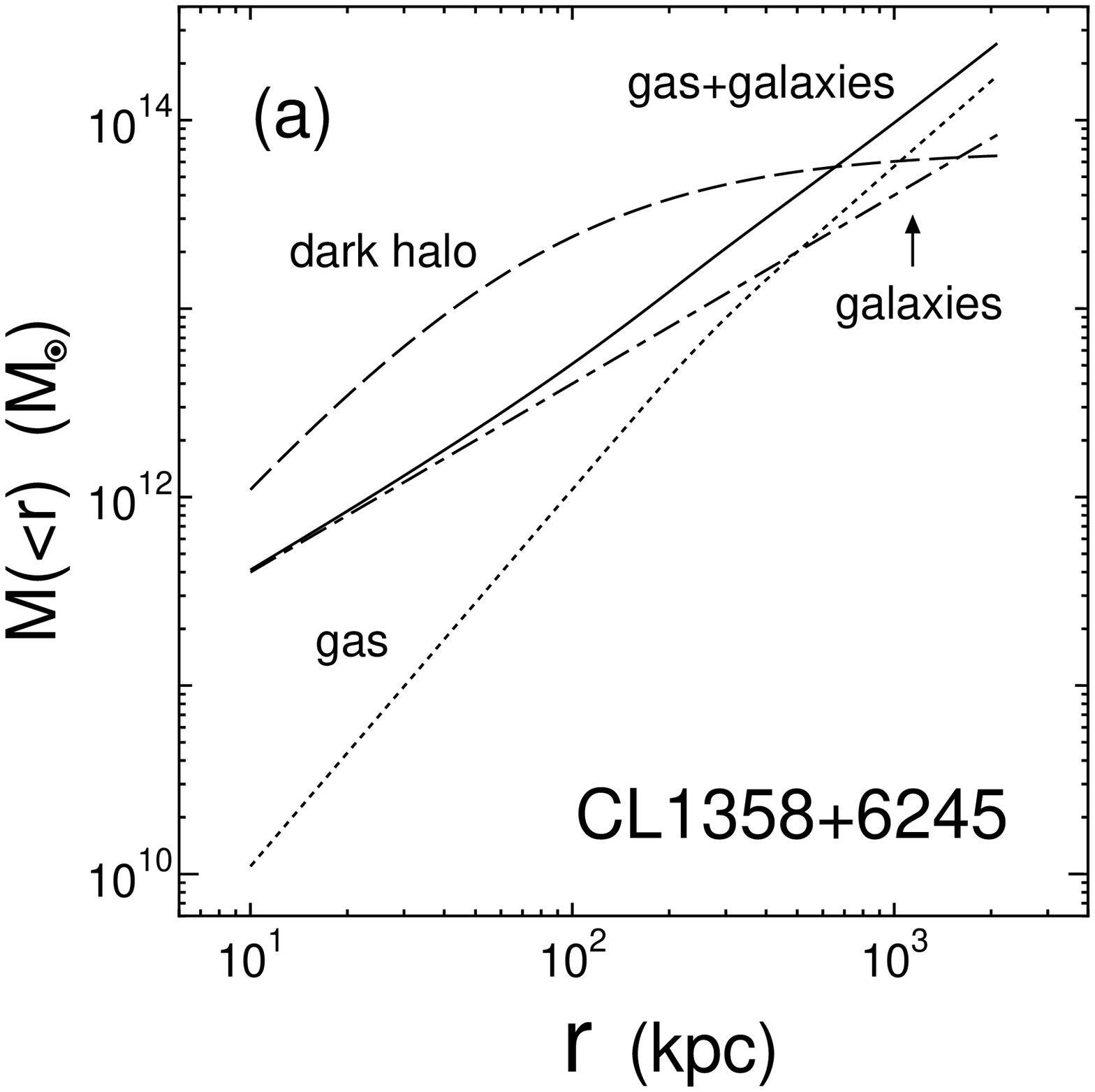}\hspace{0.5cm}\includegraphics[width=8cm]{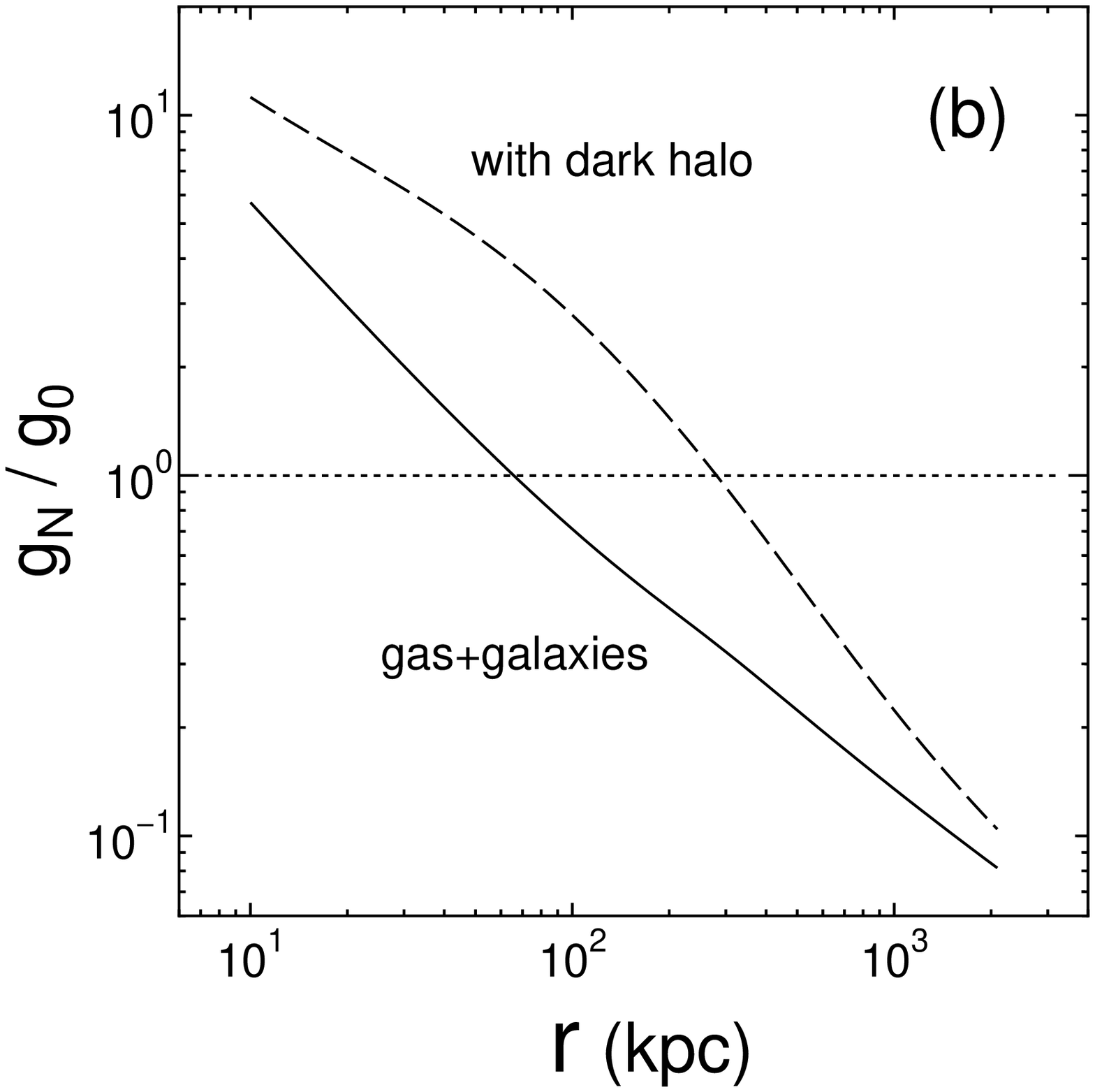}\vspace{0.5cm}
  \includegraphics[width=8cm]{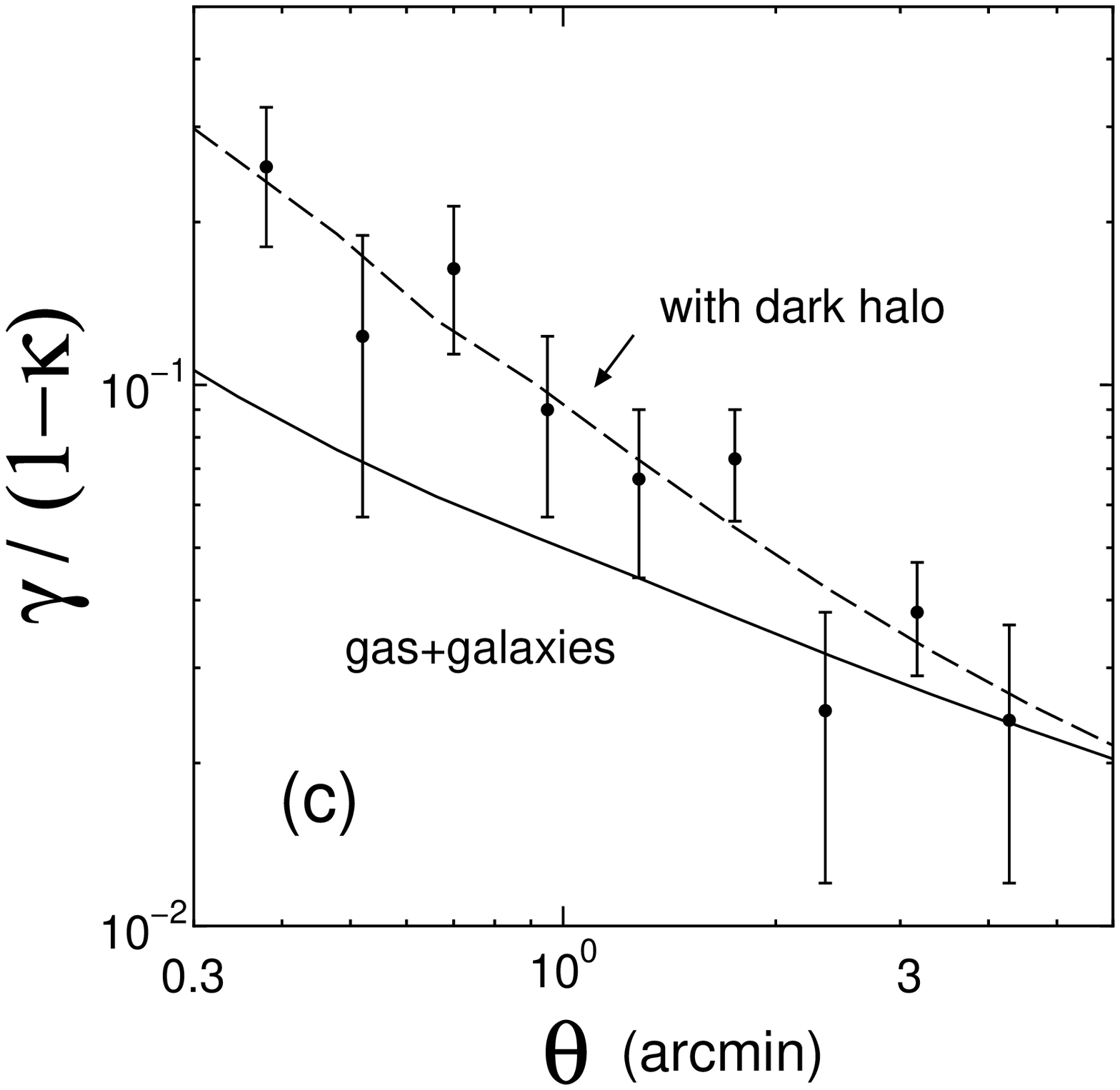}
  \caption{Same as Fig.\ref{a1689}, but for the cluster CL1358+6245.
}
\label{cl1358}
\end{figure*}

%\subsection{Other Clusters}

\subsection{SDSS clusters}

\begin{figure}
\epsscale{.80}
\plotone{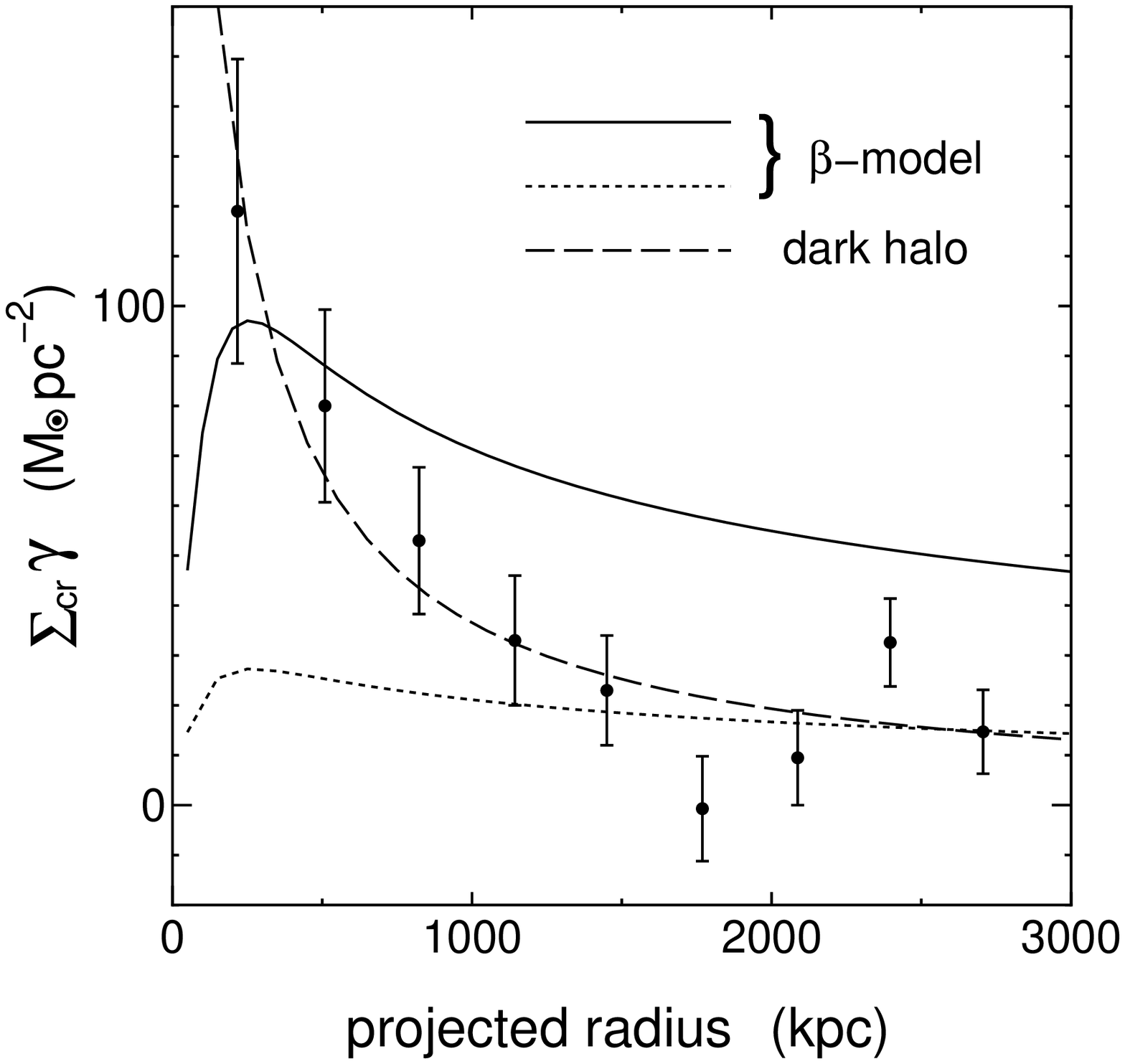}
\caption{The mean shear of $42$ SDSS clusters.
% The vertical axis is
%  $\Sigma_{cr} \gamma$ and the horizontal axis is the projected radius.
 The solid and dotted lines are the beta models, and the dashed line
 is the dark halo model.  
}
\label{sdss-cl}
\end{figure}

Sheldon et al. (2001) studied weak lensing of $42$ clusters in SDSS data.
They provided the mean shear of $42$ clusters up to the radius of $2700$ kpc
 as shown in Fig.\ref{sdss-cl}.
The vertical axis is $\Sigma_{cr} \gamma$, where
 $\Sigma_{cr}=1/(4 \pi) D_S/(D_L D_{LS})$ is the critical surface density,
 and the horizontal axis is the projected radius\footnote{The quantity
 $\Sigma_{cr} \gamma$ is related to the
 surface density of the lens in GR : $\Sigma_{cr} \gamma =
 \bar{\Sigma}(\leq R) - \Sigma(R)$.}.
Here, $\Sigma_{cr} \gamma$ does not depend on the source redshift.
The data is well fitted by a power law with index
 $-0.9 \pm 0.3$ \cite{s01}. 
On the solid and dotted lines, we consider only the gas component, 
 given by the isothermal beta model:
\beq
  \rho(r) = \rho_0 \left[ 1+\left( \frac{r}{r_c} \right)^2
     \right]^{-3 \beta/2},
\label{isb}
\eeq 
 with $\beta=0.6$, $r_c=100$ kpc, $\rho_0/\rho_{cr}=8000$ (solid line) and
 $800$ (dotted line), here $\rho_{cr}$ is the critical density at the
 present.
As shown in the figure,  the fit is poor.
This is because the slope of the shear is $-3\beta/2+1/2 = -0.4$
 for $g < g_0$ and it is flatter than the data.
%Hence including only baryonic component, the MOND can not explain the data.
The dashed line is the dark halo model (given in Eq.(\ref{hp})).
The best fitted model is $M_0=(3.8 \pm 1.1) \times 10^{13} M_\odot$
 and $r_0=75 \pm 34$kpc with $\chi^2_{min}$/dof $= 10.0/7$.
The dashed line fits the data well.

\section{Limit on Neutrino Mass}

In previous studies, several authors assumed a massive neutrino with
 a mass of $\sim 2$ eV as the
 dark matter to explain the observational data (e.g. 
 Sanders 2003; Skordis et al. 2006).
In this section, we put a constraint on its mass from the weak lensing
 of clusters.  

The neutrino oscillation experiments provide the mass differences 
 between different species
 : $\Delta m_\nu^2 \lesssim 10^{-3} \mbox{eV}^2$.
Here we consider  massive neutrinos whose masses are much heavier than
 $\Delta m_\nu$ and assume they are degenerate: they have (almost) 
the same mass, independent of species. 
Using the maximum phase space density $h^{-3}$,
 the maximum density of the neutrino dark halo is given by
 (Tremaine \& Gunn 1979; Sanders 2003; 2007) \footnote{Sanders (2007)
 recently revised his calculation and gave $\sim 1/3$ times
 smaller density  than Sanders (2003). We confirm his calculation. 
Then the minimum neutrino mass is $\sim 3^{1/4} = 1.3$ times larger.
Although Sanders (2007) is not yet published and there may be some
 ambiguities about the factor, 
we adopt his revised model.},
\beq
 \rho_{\nu,max} = 2.3 \times 10^{-5} M_\odot/\mbox{pc}^3
% \left( \frac{g}{6} \right) 
 \left( \frac{m_\nu}{2 \mbox{eV}} \right)^4
 \left( \frac{T}{\mbox{keV}} \right)^{3/2}
\label{rho_nu}
\eeq
where $T$ is the X-ray temperature : 
 $9.0 \pm 0.1$ keV for A1689 \cite{am04}, $3.5 \pm 0.2$ keV for
 CL0024 \cite{z05}, and $7.2 \pm 0.1$ keV for CL1358 \cite{abg02}.
% and $3.4 \pm 0.3$ keV for SDSS clusters \cite{s01,i02}.
%The degree of freedom is $g=6$, three flavor types and inclusion of
% anti-neutrino.
%In Eq.(\ref{rho_nu}), 
%Sanders (2007) recently revised his calculation and gave $\sim 1/3$ times
% smaller density  than Sanders (2003). We confirm his calculation. 
%Then the minimum neutrino mass is $\sim 3^{1/4} = 1.3$ times larger.
%Although Sanders (2007) is not yet published and there may be some
% ambiguities about the factor, 
%we adopt his revised model. 
% however cannot derive the previous result.  
    
For A1689, the core density of the neutrino halo is
 $\rho_c = 3 M_0/(4 \pi r_0^3)$ from Eq.(\ref{hp}).
For CL0024 and CL1358, as we noted, the core radius
 $r_0$ in the best fitting model is smaller than the inner
 most data point. 
Hence, in order to put a conservative bound, 
we use the mean density inside the second innermost data point
 $r_2$, $\rho_2 \equiv \rho(<r_2)$ $= 3 M(<r_2)/(4 \pi r_2^3)$.
Since $\rho_2 < \rho_c$, we obtain a lower bound of $m_\nu$ from
 Eq.(\ref{rho_nu}).
The results of $\rho_c$ and $\rho_2$ are shown in Table 1.
The lower row in each cluster is the case of another halo model
 $M(<r) = r^3/(r^3+r_0^3)$ instead of Eq.(\ref{hp}).
We try out this model in order to study the halo model dependence. 
For A1689 $\rho_0$ changes by a factor $\sim 5$, and 
 hence it depends on the halo profile.
However for other clusters $\rho_0$ changes slightly (less than a factor
 $2$), because 
 there are no data point near the core radius $r_0$ and
 only $M_0$ is determined through the amplitude of the
 shear.
% and the best fitting parameters of $M_0$ are almost same in two models.
%\footnote{two models are different only near $r_0$.}.
% only $M_0$ is determined through the amplitude of the
% shear and it is independent of the models.   

\begin{figure}
\epsscale{.80}
\plotone{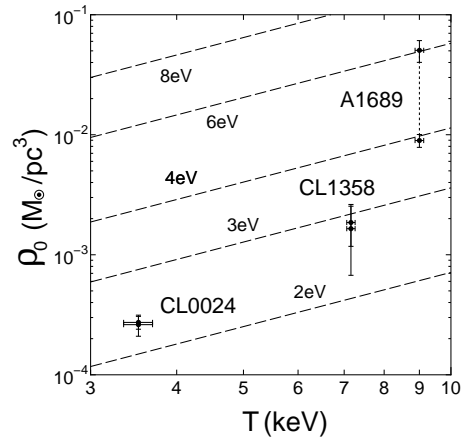}
\caption{The central density vs. the X-ray temperature.
The density $\rho_0$ is the core density for A1689, and mean density
 inside of the second innermost data point for CL0024 and CL1358.
There are two points with error bars for each cluster corresponding
 to the two density profile models in Table.1.
The vertical dotted line connecting these pairs represents the
 systematic uncertainty in the density model.
The dashed lines correspond to minimum neutrino mass in Eq.(\ref{rho_nu}).
}
\label{t-rho}
\end{figure}

In Fig.\ref{t-rho} we show the density $\rho_0$ vs. the temperature $T$
 to put a constraint on the neutrino mass.
The dashed lines correspond to neutrino mass in Eq.(\ref{rho_nu}). 
{}From the figure, the minimum neutrino mass is $2-3$ eV for CL0024 and
 CL1358.
%, and $1-2$ eV for the mean density of SDSS clusters.
The above results are consistent with the previous X-ray
 measurements \cite{s03,s07}.
Since the current limit is $m_\nu < 2$ eV from tritium $\beta$
 decay\footnote{Particle
 Data Group Home Page : http://pdg.lbl.gov/},
 these values are comparable to or larger than this limit.

%We note that we use the mean density inside of the
% second innermost data point $\rho_2$ which is lower than the core density.
The mean density inside the innermost, not second innermost, data point
 is much higher than $\rho_2$.
% : for CL0024 $24 (25)$ times and for CL1358
% $6.5 (1.5)$ times larger than the results in the upper (lower) rows
% of Table.1.
The neutrino masses are $5-6$ eV for CL0024 and $3-5$ eV for
 CL1358 in this case.
Hence there is an ambiguity about the definition of the central
 density for these clusters.

The core density of A1689 is highest and
 the minimum neutrino mass reaches $4-6$ eV.
To check our result, we compare the core density with the previous
 studies in GR.
Since $g_N \sim 10 \times g_0$ for $r < r_0$ from Fig.\ref{a1689}(b), 
 GR is valid near the core.
Halkola et al. (2006) gave central mass distribution by
 analyzing $107$ multiple images of $32$ background galaxies.
Their mass distribution is consistent with Broadhurst et al. (2005b)
 (see Fig.17 of Halkola et al. (2006)).
From a velocity dispersion of $1450$ km/s
 and the core radius $77$ kpc, the core density is
 $\rho_c=0.01 M_\odot/\mbox{pc}^3$.
This is roughly consistent with our result.

%Recently Nakajima \& Morikawa (2007) studied the similar topic.
% solved the hydrodynamical equilibrium
% equation with degeneracy pressure of fermionic dark matter.
%They conclude $2-4$ eV for $g=1-4$.

%Sanders (2003) derived a core radius of neutrino virialized halo,
% $r_\nu \gtrsim 180 \mbox{kpc}$ $(m_\nu/4 \mbox{eV})^{-2}$
% $(T/\mbox{keV})^{-1/4}$.
%Our results satisfy this condition.

Allen (1998) suggests that the lensing core mass is generally
 a few times larger
 than X-ray core mass for non-cooling flow clusters
 (see also Clowe \& Schneider 2001 in the case for A1689).
Some ideas are proposed to explain the discrepancies :
 clusters are not in dynamical equilibrium,
 non-thermal pressure such as turbulent and magnetic pressure
 plays an important role,
 elongation of the cluster or substructures along a line-of-sight
 (e.g. Hattori, Kneib \& Makino 1999). 
In fact, Lokas et al. (2006) show that A1689 is
 surrounded by a few substructures aligned along a line-of-sight,
 by studying the galaxy kinematics. 
CL0024 has a second mass clump which is separated at $3^\prime$
 from the center and has $30 \%$ of total cluster mass
 \cite{k03}.
Jee et al. (2007) recently suggest that CL0024 would be the merging
 cluster in line-of-sight direction.
These systematics would affect our results and change the 
 neutrino mass limit by factor
 (since $m_\nu$ is not very sensitive to $\rho_c$,
 $m_\nu \propto \rho_c^{1/4}$ from Eq.(\ref{rho_nu})).

\begin{deluxetable}{cccc}  
\tablecaption{Best fitting model for dark halo profile and central density}
\tablewidth{0pt}
\tablehead{
 & $M_0 ~(M_\odot)$ & $r_0 ~(\mbox{kpc})$ 
 & $\rho_0 ~(M_\odot/\mbox{pc}^3)$
}
\startdata
 A1689 & $(1.1 \pm 0.1) \times 10^{15}$ & $174 \pm 11$
 & $(5.1 \pm 1.0) \times 10^{-2}$ \\
 & $(5.1 \pm 0.3) \times 10^{14}$  & $239 \pm 9$
 & $(9.0 \pm 1.1) \times 10^{-3}$ \\
 \hline 
 CL0024 & $(2.0 \pm 0.4) \times 10^{14}$ & $42 \pm 15$
 & $(2.6 \pm 0.6) \times 10^{-4}$ \\
 & $(1.7 \pm 0.2) \times 10^{14}$ & $147 \pm 19$
 & $(2.8 \pm 0.3) \times 10^{-4}$ \\
 \hline
 CL1358 & $(6.9 \pm 2.8) \times 10^{13}$ & $69 \pm 31$
 & $(1.7 \pm 1.0) \times 10^{-3}$   \\
 & $(4.0 \pm 1.1) \times 10^{13}$ & $127 \pm 27$ &
 $(1.9 \pm 0.7) \times 10^{-3}$ 
%\hline
% SDSS & $(3.8 \pm 1.1) \times 10^{13}$ & $75 \pm 34$
% & $(4.6 \pm 1.3) \times 10^{-5}$   \\
% & $(3.1 \pm 0.6) \times 10^{13}$ & $206 \pm $ 34&
% $(5.2 \pm 1.1) \times 10^{-5}$
\enddata
\tablecomments{
 The best fitting parameters $M_0$ and $r_0$.
 The density $\rho_0$ is the core density $\rho_c$ for A1689
 and the mean density inside the second innermost data point
 $\rho_2$ for the others.
 The upper row is the case of the halo profile $M(<r)=M_0 r^3/(r+r_0)^3$
 given in Eq.(\ref{hp}), while the lower row is $M(<r)=M_0 r^3/(r^3+r_0^3)$.
% We note that $\rho_2 < \rho_c$ is more conservative density 
% For CL0024 and CL1358, the core radius is smaller than 
% the inner most data point $r_{min}$ ($140$ kpc for CL0024 and $110$ kpc
% for CL1358). Hence it only means that the core radius
% is smaller than $r_{min}$.
}
\label{table1}
\end{deluxetable}

\section{Results by Other Interpolation Functions}

So far, we used a standard interpolation function
 $\tilde{\mu}(x)=$ $x/\sqrt{1+x^2}$ alone. 
However the standard $\tilde{\mu}$ is not consistent with 
 T$e$V$e$S \cite{b04}.
In this section, we also examine other interpolation functions,
 Bekenstein's toy model in T$e$V$e$S
 $\tilde{\mu}(x)=$ $4x/(1+\sqrt{1+4x})^2$
 and simple model $\tilde{\mu}(x)=$ $x/(1+x)$ in Famaey \& Binney (2005),
 in order to study the robustness of our results.
Since the clusters have $g_N \approx g_0$ from
 Figs.\ref{a1689}-\ref{cl1358} (b), our conclusions may depend on
 the choice of $\tilde{\mu}$.

\begin{figure}
\epsscale{.80}
\plotone{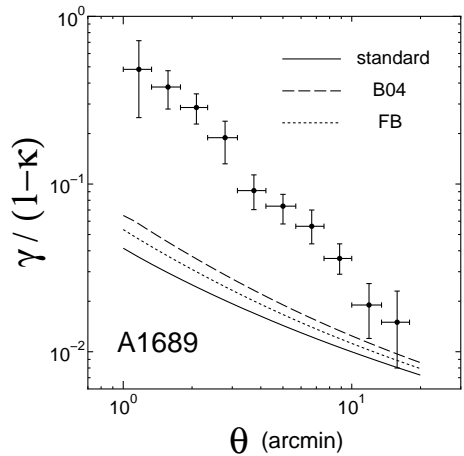}
\caption{The MOND predications for A1689 for the three interpolation
 functions.
The solid line is the standard model $\tilde{\mu}(x)=$ $x/\sqrt{1+x^2}$,
 the dashed line is the Bekenstein's toy model (B04) 
 $\tilde{\mu}(x)=$ $4x/(1+\sqrt{1+4x})^2$, and the dotted line
 is the Famaey \& Binney's model (FB) $\tilde{\mu}(x)=$ $x/(1+x)$.}
\label{a16v}
\end{figure}

In Fig.\ref{a16v}, we show the MOND predication for A1689 for the
 three types of $\tilde{\mu}(x)$.
The solid line is the standard model,
 the dashed line is the Bekenstein's toy model (B04),
 and the dotted line is the Famaey \& Binney's model (FB).
The standard model predicts the lowest value,
 because $\tilde{\mu}(x)$ is highest at $x \sim 1$.
 B04 and FB show several times $10 \%$ higher values
 than the standard model.
%, the discrepancy can not be solved.
%We also show the best fitting halo models in Table 1.
%Although the quantitative results depend on $\tilde{\mu}(x)$, 
% the dark halo is necessary for any models.

The minimum neutrino mass only $\sim 10 \%$ changes 
% (see Table.1 of our manuscript ver.3 in astro-ph arXiv)
 and the choice of $\tilde{\mu}$ is not crucial.

\section{Conclusion}

We have studied the weak lensing of galaxy clusters in MOND.
We calculate the shears and the convergences of the background galaxies
 for three clusters (A1689, CL0024, CL1358) and
 the mean profile of 42 SDSS clusters,
 and compare them with the observational data.
It turns out that the MOND cannot explain the data irrespective of $g_0$
 unless a dark matter halo is added.
We also examine the three types of interpolation function, but the
 conclusion does not change. 
The above results are consistent with those of previous studies (e.g. 
 Aguirre, et al. 2001; Sanders 2003).
If the dark halo is composed of massive neutrinos, its minimum mass 
 is $4-6$ eV for A1689 and $2-3$ eV for CL0024 and CL1358. 
However our results still depends on the dark halo model and
 inner data points.
Even for A1689, the system with the most constraining data, the
 systematic is more important than random errors as shown in
 Fig.\ref{t-rho}.
In addition, there are some systematic uncertainties such as an elongation
 of cluster along line-of-sight.   
These effects would reduce the minimum mass by a factor of $\sim 2$.
In the low acceleration region ($g_N \lesssim g_0$), the external
 gravitational field becomes important and would affect the outer shear
 profile \cite{bm84,wu07}. 

In conclusion,
 the more careful study is necessary to put a stringent constraint,
% (error is less than $\sim 10 \%$), 
 for example, a combination of the weak and strong lensing,
 X-ray gas and galaxy dynamics.
%These values are relatively high in comparison with the previous results
% determined by X-ray measurements \cite{s03}. 
However, it is beyond the scope of this paper and we will study
 as an future work.
Even so, at present, 
we find that there is some tension between the lower bound of neutrino mass 
in neutrino dark halo model in MOND and the upper bound by experiments.  
%Even with massive neutrinos, it is still  difficult to form the small
% core ($<200$ kpc) in the neutrino halo model.

\acknowledgments
We would like to thank Takashi Hamana for useful comments and discussions. 
We also thank the anonymous referee for helpful comments to improve the
 manuscript. 
TC was supported in part by a Grant-in-Aid for Scientific Research (No.17204018) 
from the Japan Society for the Promotion of Science and in part by Nihon University.

\clearpage

\end{document}